\documentclass[iop,revtex4-1]{emulateapj}

\usepackage{epsfig}
\usepackage{epstopdf}
\usepackage{graphics}
\newcommand{\RMxAA}{Revista Mexicana de Astronomia y Astrofisica}

\slugcomment{To appear in the Astrophysical Journal.}

\shorttitle{An ionized jet from the massive protostar in G11.11$-$0.12P1}
\shortauthors{Rosero et al.}

\begin{document}

\title{Weak and Compact Radio Emission in Early Massive Star Formation Regions: An Ionized Jet Toward G11.11$-$0.12P1}

\bigskip\bigskip
\newcommand{\nraoblurb}{The National Radio Astronomy Observatory is
a facility of the National Science Foundation operated under cooperative
agreement by Associated Universities, Inc.}

\author{V. Rosero$^{1}$, P. Hofner$^{1,}$\altaffilmark{\dag}, M. McCoy$^{1}$, S. Kurtz$^{2}$,
K. M. Menten$^{3}$, F. Wyrowski$^{3}$, E. D. Araya$^{4}$, L. Loinard$^{2}$, C. Carrasco-Gonz\'alez$^{2}$, L. F. Rodr\'iguez$^{2}$, R. Cesaroni$^{5}$, S. P. Ellingsen$^{6}$}

\affil{$^1$Physics Department, New Mexico Tech, 801 Leroy Pl., Socorro, NM 87801, USA}

\affil{$^2$Centro de Radioastronom{\'\i}a y Astrof{\'\i}sica, 
Universidad Nacional Aut\'onoma de M\'exico, Morelia 58090, M\'exico}

\affil{$^3$Max-Planck-Institute f\"{u}r Radioastronomie, Auf dem H\"{u}gel 69, 53121 Bonn, Germany}

\affil{$^4$Physics Department, Western Illinois University, 1 University Circle, Macomb, IL 61455, USA}

\affil{$^5$INAF, Osservatorio Astrofisico di Arcetri, Largo E. Fermi 5, 50125 Firenze, Italy}

\affil{$^6$School of  Physical Sciences, University of Tasmania, Private Bag 37, Hobart, Tasmania 7001, Australia}

\altaffiltext{\dag}{Adjunct Astronomer at the National Radio Astronomy Observatory.}

\begin{abstract}

We  report 1.3$\,$cm and $6\,$cm continuum observations toward  the massive proto-stellar candidate G11.11$-$0.12P1 using the Karl G. Jansky Very Large Array (VLA).
We  detect  a string of four unresolved radio continuum  sources coincident with the mid-IR source in G11P1. The continuum sources  have positive spectral indices consistent with a
thermal (free-free)  ionized jet. The most likely origin of the ionized gas are shocks due to the interaction of a stellar wind with the surrounding high-density material. 
We also present  NIR United Kingdom Infrared Telescope (UKIRT) archival data which show an extended structure detected only at K-band ($2.2~\mu$m), which is oriented perpendicular
to the jet, and that may be scattered light from a circumstellar disk around the massive protostar. Our observations plus the UKIRT archival data thus provide new evidence that a disk/jet system is present in the massive protostellar candidate located in the G11.11$-$0.12P1 core.

\end{abstract}

\keywords{ISM: individual (G11.11$-$0.12P1)  -- ISM: jets -- stars: formation}

\section{Introduction}
The role of jets in massive star formation is not yet fully understood. Unlike  their low-mass counterparts,
the current sample of known massive young stellar objects (MYSOs) associated
with collimated jets is very small (see \citealt{2010ApJ...725..734G} for a summary).
MYSOs are difficult to detect since they are located at large distances, 
with a tendency to form in complicated cluster environments, and evolve on a much shorter evolutionary timescale
compared to  low-mass stars. It is important, therefore, to identify more candidates
of massive stars in early evolutionary stages to ascertain whether jets are present, and if so, to study their role during the formation process.
Infrared dark clouds (IRDCs) are potentially a good place to find molecular cores which might harbor the earliest
stages of massive star formation (e.g. \citealt{2000ApJ...543L.157C}). IRDCs are cold (T$<$ 25 K), high column density  
($\sim$~10$^{23}$ $-$ 10$^{25}$ cm$^{-2}$) molecular condensations, with high gas densities ($>10^{5}$ cm$^{-3}$)
and a large amount of extinction (A$_{\textrm{v}}\sim $ 200 mag, \citealt{2014ApJ...782L..30B}), which causes them to appear as dark silhouettes against the Galactic mid-infrared background \citep{1998ApJ...508..721C, 2005IAUS..227...23M, 2006ApJ...641..389R}.

\begin{figure*}
\centering

\includegraphics[scale=0.25]{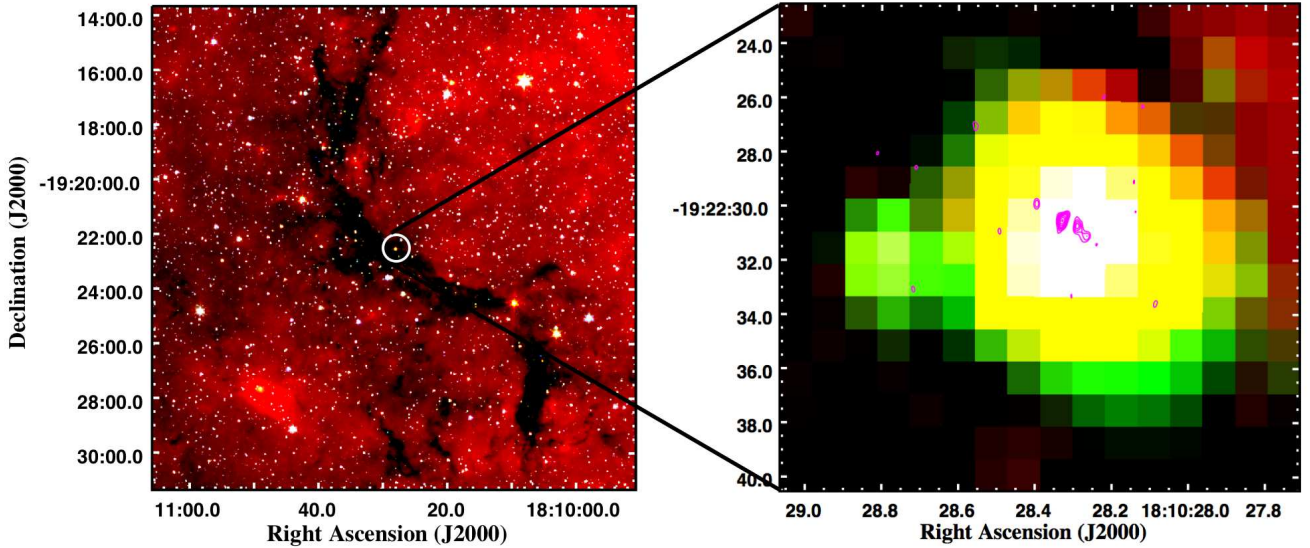}

\caption{Left: \emph{Spitzer} IRAC GLIMPSE three-color (3.6-blue, 4.5-green and 8.0-red $\mu m$)
image of  the filamentary IRDC G11.11$-$0.12. Right: Three color image of the center core 
G11P1 overlayed with VLA 6 cm continuum emission contours.
The 6 cm contour levels are (3, 4, 5, 6, 7, 9, 11, 13)  $\times$  5$\,\mu$Jy . 
The right panel shows 4.5 $\mu$m excess emission towards  G11P1.}\label{f1}
\end{figure*}

G11.11$-$0.12P1 (hereafter G11P1) is a compact  dust continuum  source located in the filamentary IRDC
G11.11$-$0.12 at a kinematic distance of 3.6 kpc (\citealt{1998ApJ...508..721C, 2000ApJ...543L.157C}; \citealt{2003ApJ...588L..37J}). 
 Figure~\ref{f1} shows a \emph{Spitzer} IRAC GLIMPSE three-color image of the G11.11$-$0.12 IRDC. The right panel shows the G11P1 core, along with our VLA 6 cm image (see below). 
Several indicators  show that G11P1 is an active star forming
region: {\it i)} compact sub-mm dust continuum (450 $\mu$m and 850 $\mu$m; \citealt{2000ApJ...543L.157C}), {\it ii)}  point-like mid-IR emission (8 $\mu$m, \citealt{2000ApJ...543L.157C}; 24 $\mu$m, \citealt{2011A&A...529A.161G}),  {\it iii)}  H$_{2}$O and class II CH$_{3}$OH maser emission (\citealt{2006A&A...447..929P}, hereafter P06), and {\it iv)} outflow indicators such as   4.5 $\mu$m excess emission \citep{2008AJ....136.2391C}.

The luminosity of G11P1 estimated from a  spectral energy distribution (SED) model is
$\sim$1200 L$_{\odot}$ (P06). The SED peaks in the far-IR
but also has a mid-IR component that P06 attribute to an accretion disk.
\citet{2010A&A...518L..95H} observed G11P1  
using \emph{Herschel} PACS  at
70 $\mu$m, 100 $\mu$m and 160 $\mu$m. Their SED model  suggests a dust temperature of 24~K and a core mass of 240 M$_{\odot}$, corresponding to a luminosity
of 1346~L$_{\odot}$, the largest of the sources in the  G11.11$-$0.12 IRDC. G11P1 has also been detected in the dense gas tracers  NH$_{3}$ (P06), C$^{34}$S 
as well as in several thermally excited (i.e. non-maser) transitions of  CH$_3$OH \citep{2011A&A...529A.161G}.

P06 detected a strong (22~Jy for the brightest peak) class II methanol maser at 6.7 GHz in G11P1 using the Australian Telescope Compact Array (ATCA). They reported a velocity structure with a 
linear trend  which they interpreted as a  disk around a highly embedded massive protostar. In addition, a  2MASS NIR emission structure detected 2$^{\prime\prime}$ from the maser 
supports the circumstellar disk scenario (P06; more discussion in this
regard is given in \S \ref{near_ir}). P06 also detected a weak ($\sim$0.3 Jy) water maser at 22.2 GHz using the VLA in the D
configuration. In this case the velocity structure of the maser spot is not spatially resolved; the water maser is slightly offset ($\sim$ 1$^{\prime\prime}$) from the methanol maser position.
Both maser species are indicators of the earliest stages of massive star formation, and in particular the 6.7 GHz CH$_3$OH maser has only been found in regions
were massive stars form \citep{2003A&A...403.1095M, 2013MNRAS.435..524B}.

In a recent paper \citet{2014MNRAS.439.3275W} presented SMA and VLA continuum and molecular line observations toward G$11.11-0.12$, which showed that the P1 core
contains 6 condensations with masses in excess of the thermal Jeans masses. They also reported the discovery of an East-West outflow, which is most clearly seen in the SiO(5--4) line.

None of the previous observations were sufficiently sensitive to detect the cm continuum 
towards G11P1. Our new high sensitivity VLA observations presented in this paper show the presence of
cm continuum sources associated with the mid-IR point source.
All of the features discussed above make G11P1 a strong candidate for an  embedded
massive young stellar object (MYSO) in an early stage of formation, and likely hosting an outflow/disk system.

In this paper, we present sensitive sub-arcsecond resolution continuum observations of G11P1 at
6 cm and 1.3 cm using the Karl G. Jansky Very Large Array (VLA)\footnote{The National Radio Astronomy Observatory is a facility of the National Science Foundation operated under cooperative agreement by Associated Universities, Inc.}.
These observations were made as part of a larger survey to search for weak, compact radio emission in young, high-mass
star forming regions. The results of the survey will be presented elsewhere (Rosero et al. in preparation); here we present the results
for G11P1. 
We describe our VLA observations and data reduction in \S\,\ref{data_red},
in \S\,\ref{results} we present our observational results of the radio continuum data, in \S\,\ref{analysis} we present
an analysis of our cm detections and of the NIR emission,
in \S\,\ref{discussion} we discuss the nature of the massive protostar in G11P1,
and in \S\,\ref{conclusions} we  summarize our findings.

%%%%%%%Figure 1 
%%%%%%%%%%%%%%%%%%%%%%%%%%%%%%%%%%%%%%

%%%%%%%%%%%%%%%%%%%%%%%%%%%%%%%%%%%%%%

\section{Observations and Data Reduction}\label{data_red}
VLA  continuum observations  (project code 10B-124) 
at 6 and 1.3 cm were obtained  for the core 
region G11P1. The pointing center was RA(J2000)=$18^{h}10^{m}28{\rlap.}^{s}40$,
Dec(J2000)$=-$$19^{\circ}22^{\prime}29{\rlap.}^{\prime\prime}0 $.
The observations where made in different 
configurations --- A-configuration for 6 cm and B-configuration for 1.3 cm ---
to obtain similar angular resolution at the different 
frequencies.
Following is a detailed description of the observations.

\subsection{6 cm Observations}
The observations were made in the A-configuration  on 
2011 July 27 covering two 1~GHz wide basebands
  centered at 4.9 and 7.4 GHz, respectively.
Each band was divided into 8$\times$128 MHz
spectral windows (SPWs).
Therefore, the data were recorded in 16 unique SPWs, each of these
with 64 channels (resolution $=2$ MHz), i.e.  a total bandwidth of 2048 MHz.
The SPWs were configured to avoid the strong methanol maser emission at 6.7 GHz.
For flux calibration we observed 3C286 and
the phase calibrator was J1820$-$2528. 
Alternating observations between G11P1 and the phase calibrator
were made with on-source times of 900 s and 180 s, respectively.
The total observing time was $\sim\,$1~hr, of which $\sim\,$40 minutes were 
on-source. All 27 antennas were available  after flagging.\\

The data were processed  using NRAO's Common Astronomy Software 
Applications (CASA\footnote{http://casa.nrao.edu}).  
Eight channels at the edges of each baseband  
were flagged due to  substantial roll-off (and therefore loss of sensitivity). 
In addition, a large amount of radio frequency interference (RFI) was flagged throughout the observing band (approximately 20$\%$ of the total data). 
The bandpass solution 
was formed using 3C286. This solution was  applied when solving for the complex gains.
The flux density  for 3C286 was adopted from the Perley--Butler 2010  flux calibration standards, and the derived flux density
for the phase calibrator at 6.086 GHz was  1.026 $\pm$ 0.002 Jy with spectral index of $-$0.29.
The gain solutions were then applied to the target source G11P1.
The images were made using Briggs ${\tt ROBUST}=0.5$ weighting. Owing to the low S/N of the detections
($<$ 20), no self-calibration was attempted. 

As a consistency check, and to ensure the absence of line contamination or RFI, we imaged and 
inspected each SPW separately. Each 1~GHz baseband was imaged separately to provide a better estimate 
of spectral index. Finally, a combined image was made, including all data from both basebands.
The   synthesized 
beam of this combined image is $0.49^{\prime\prime} \times 0.27^{\prime\prime}$,
position angle PA $=172^{\circ}$, and rms noise  $\sim 5\,\mu$Jy beam$^{-1}$. \\

\subsection{1.3 cm Observations}
The observations were made in the B-configuration  on 
2011 March 20  covering two  1 GHz wide bands
centered at 21 and 25.5 GHz. 
Each band was divided into 8$\times$128 MHz
SPWs. The SPWs were configured to avoid the strong  water maser emission at 22~GHz.
The same number of  SPWs and
channels  were used as in the  6 cm observations.
For flux calibration we observed 3C286 and
the phase calibrator was J1820$-$2528. 
Alternating observations between the target and the phase calibrator
source were made with   times of 270 and 90 s, respectively.
The  total on-source time  was  $\sim\,$42 minutes. After flagging,
only 23 antennas were available. 
Pointing corrections were obtained separately and applied  during the observations.\\

The data reduction was done in the same fashion as for the 6 cm observations.  
The flux density  for 3C286 was adopted from the Perley--Butler 2010  flux calibration standards, and the derived flux density
for the phase calibrator at 23.186 GHz was  0.91 $\pm$ 0.01 Jy with spectral index of $-$0.57.
The images were made using natural  weighting. 
Opacity corrections were applied during calibration.

The absence of line contamination and RFI was confirmed by imaging each SPW separately.
As at 6 cm, we imaged each baseband individually (for spectral index) and together 
(for morphology and improved S/N).
The  synthesized 
beam of the combined map is $0.75^{\prime\prime} \times 0.28^{\prime\prime}$,
PA $=146.6^{\circ}$, and rms noise  $\sim8\,\mu$Jy beam$^{-1}$.\\

%%%%%%%%%%%%%%%%%%%%%%%%%%%%%%%%%%%%%%

\begin{figure*}
\centering$
\begin{tabular}{cc}
\hspace{-1. cm}\includegraphics[scale=0.5]{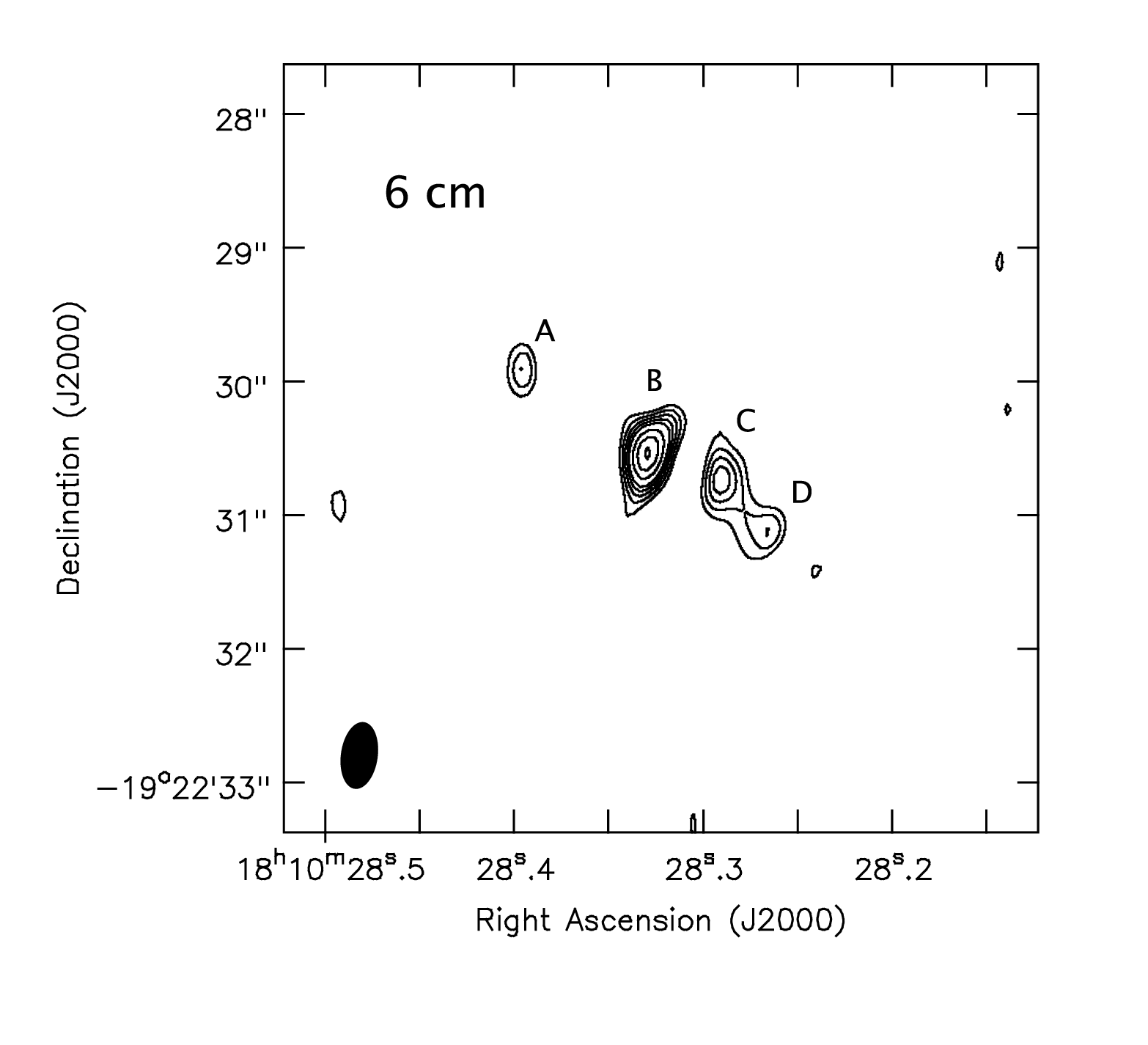}&\hspace{-1.0 cm}
\includegraphics[scale=0.5]{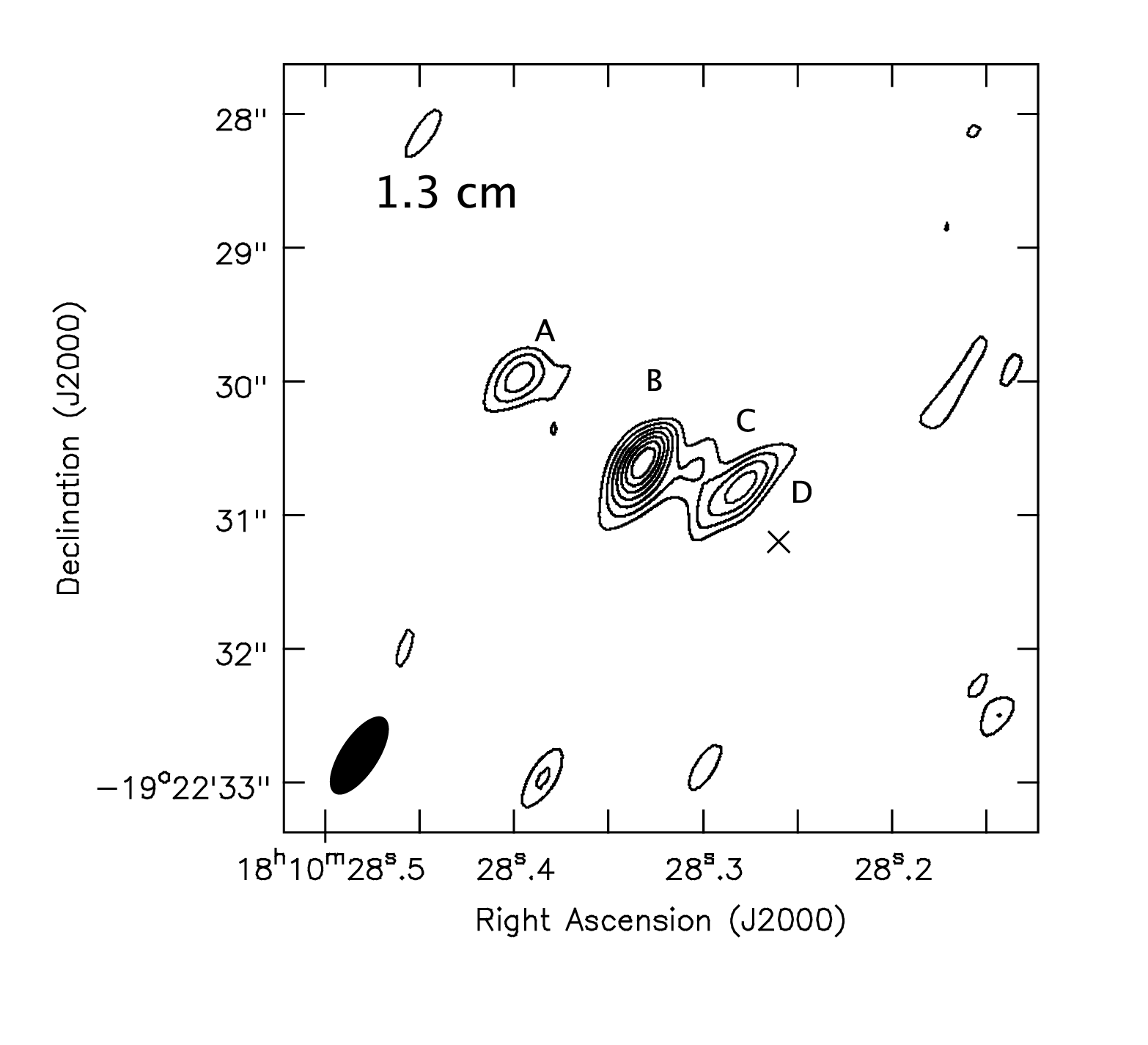}
\end{tabular}$

\caption{VLA  contour plots of the 6 and 1.3 cm images toward G11.11$-$0.12P1. 
The synthesized beam  is shown in the lower left corner,  $0.5^{\prime\prime} \times 0.3^{\prime\prime}$ for the 6 cm and  $0.7^{\prime\prime} \times 0.3^{\prime\prime}$ for 1.3 cm
combined maps. 
The 6 cm contour levels are (3, 4, 5, 6, 7, 9, 11, 13)  $\times$  5$\,\mu$Jy beam$^{-1}$
and the 1.3 cm contour levels are (2.5, 3.5, 4.5, 5.5, 6.5, 7.5, 8.5) $\times$  8$\,\mu$Jy beam$^{-1}$.
The components are labeled alphabetically
from east to west. The $\times$ symbol in the 1.3 cm map represents the location
of the D component detected in the 6 cm map.}\label{f2}
\end{figure*}
%%%%%%%%%%%%%%%%%%%%%%%%%%%%%%%%%%%%%%

\section{Results}\label{results}
We detected radio continuum emission at all observed frequencies. The emission is clearly associated 
with G11P1 (see Figure~1, right panel). In Figure \ref{f2} we show contour plots of G11P1 at 6 and 1.3 cm. 
Four and three components are detected  in the 6 cm and 
1.3 cm maps, respectively. 
As indicated in  figure \ref{f2}, we refer to these components, from 
east to west,  as A, B (bright central source), C, and D.  
The components lie in a linear structure with a PA 
of $\sim55^{\circ}$. The outermost sources 
(A and D) are separated by an angular distance of $\sim2.5^{\prime \prime}$ 
(9000 AU at the distance of 3.6 kpc). Component D is not detected at 1.3 cm.\\

Table \ref{tab2} lists the peak positions and flux densities of components A -- D, as determined by gaussian fits using the IMFIT CASA routine.
The astrometric accuracy of the VLA is better than 0.1$^{\prime\prime}$. Components A and B have consistent peak positions, but component C appears slightly offset between $1.3$ and
$6\,$cm. Such an offset can occur if the continuum optical depth in the source varies strongly between the two observing frequencies; however since
the offset is smaller than our resolution element, for simplicity we will treat the two emission peaks as a single source,
component C.
The radio components are mostly unresolved, implying
upper limits on the size of the emitting regions of about $1800\,$AU.  
This small size of the emitting regions is also reflected in the low measured brightness temperatures within a synthesized beam,
which are $\leq 17\,$K for all components.

\begin{figure*}
\centering
\includegraphics[scale=0.8]{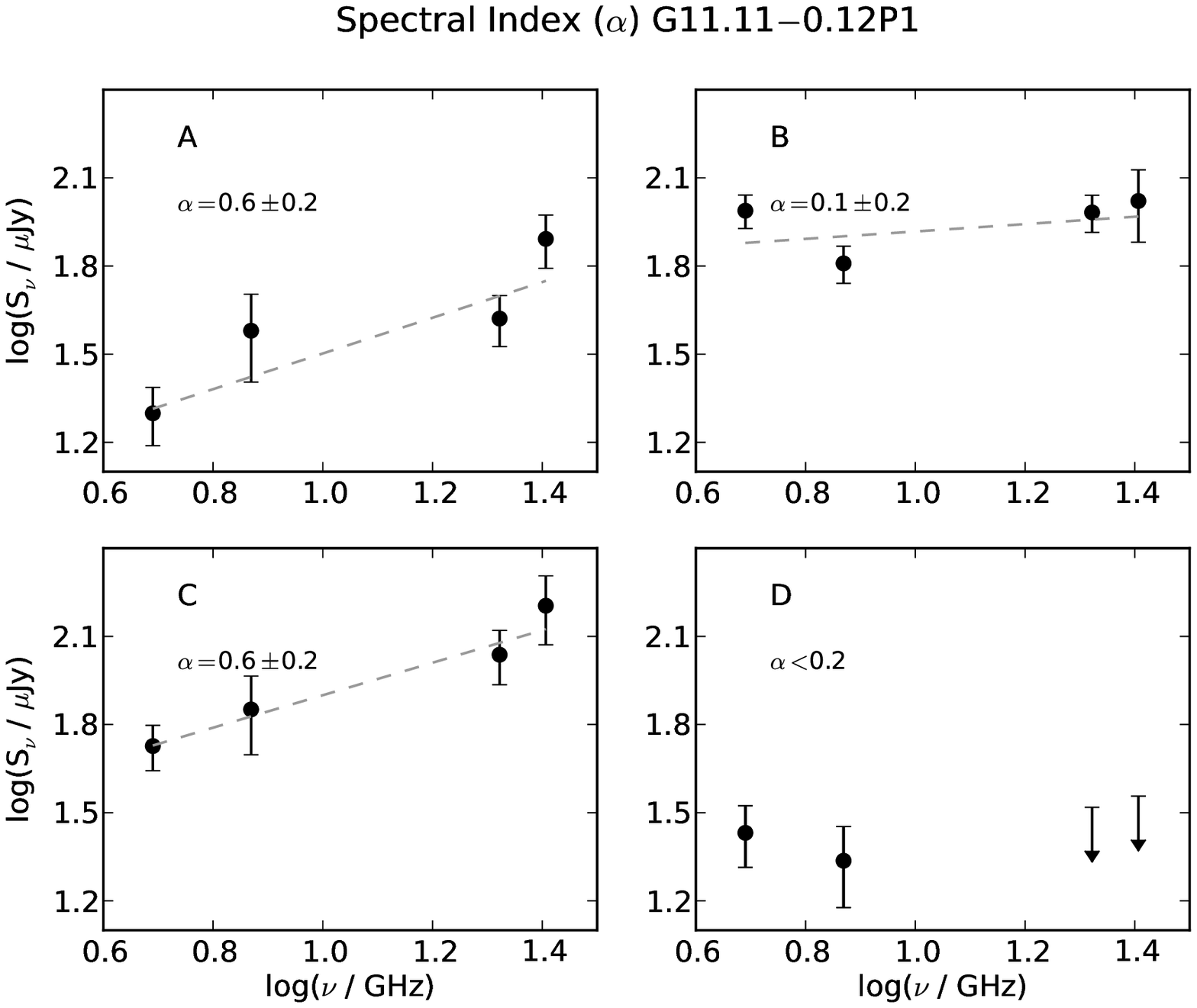}

\caption{Flux density as a function of frequency for each detected component toward 
G11.11$-$0.12P1. Error bars are the uncertainties as reported by IMFIT added in quadrature with
an assumed 10$\%$ error in calibration. The dashed lines are the best fit to the data from a power law of the form $S_{\nu} \propto \nu^{\alpha}$.}\label{f3}
\end{figure*}

Figure \ref{f3} shows the fluxes and  power law fits of the form $S_{\nu} \propto \nu^{\alpha}$,
where $\alpha$ is the spectral index and $\nu$ is the frequency for each detected 
component in G11P1.
Components A and C have a rising spectral index indicative of thermal emission from ionized gas
from a stratified medium, and component B has a flat behavior which is consistent with emission
from  optically thin ionized gas. For component D the 1.3 cm detection limits together
with the $6\,$cm data are consistent with a flat spectrum, but a falling spectral index, as
expected for non-thermal emission, cannot be excluded.

Our measured flux densities are consistent with the non-detection at $8.64\,$GHz by P06.
Also, extrapolating our flux densities to 3 mm  with a spectral index of 0.6 results in values far below 
the 12 mJy reported by P06, 
thus confirming their result that the 3 mm emission is likely due to dust.

%%%%%%%%%%%%%%%Table
\renewcommand{\thefootnote}{\alph{footnote}} 
\begin{deluxetable*}{c c c c c c}

%\begin{minipage}{\textwidth}

\centering
\tabletypesize{\scriptsize}
\tablecaption{Continuum Parameters of the Components Observed toward G11.11$-$0.12P1\label{tab2}}
\tablewidth{0pt}

\tablehead{
\colhead{Component}   &
\colhead{$\nu_{c}$} & 
\colhead{RA} & 
\colhead{Dec} & 
\colhead{S$_{\nu}$}   &
\colhead{Spectral}  \\[2pt]
\colhead{G11.11-P1} & 
\colhead{(GHz)} &
\colhead{(J2000)}& 
\colhead{(J2000)} & 
\colhead{($\mu$Jy)} & 
\colhead{index}\\[-5pt]\\}

\startdata

  A 		& 4.9   & 18 10 28.39 & $-$19 22 29.8 & 20.0(4.0)     		&  $+$0.6(0.2) \\[3pt]  
		& 7.4   & 18 10 28.40 & $-$19 22 30.1 & 38.0(12.0)  			&          \\[3pt]  
		& 20.9 & 18 10 28.40 & $-$19 22 29.9 & 41.8(7.1)   			&          \\[3pt]  
		& 25.5 & 18 10 28.40 & $-$19 22 30.1 & 78.0(14.0)  			&          \\[9pt]

 B 		& 4.9  & 18 10 28.33 & $-$19 22 30.5 & 97.3(8.1)      	 	 &  $+$0.1(0.2) \\[3pt]  
		& 7.4  & 18 10 28.33 & $-$19 22 30.5 & 64.4(6.7)       	  	 & 			 \\[3pt]  
		& 20.9 &18 10 28.33 & $-$19 22 30.7 & 96.0(10.0)    		 &                          \\[3pt]  
		& 25.5 &18 10 28.34 & $-$19 22 30.7 & 105.0(27.0)  		 &                          \\[9pt]

C 		& 4.9  & 18 10 28.29 & $-$19 22 30.7 & 53.3(7.7)\tablenotemark{b}      	  	 & $+$0.6(0.2)	 \\[3pt]  
	         & 7.4  & 18 10 28.29 & $-$19 22 30.8 & 71.0(20.0)\tablenotemark{b}	 		 & 	    		 \\[3pt]  
		& 20.9 &18 10 28.29 & $-$19 22 30.9 & 109.0(20.0)\tablenotemark{b}	          &         		 \\[3pt]  
		& 25.5 &18 10 28.28& $-$19 22 30.7 & 160.0(39.0)\tablenotemark{b}  		&        		  \\[9pt]  

D 		& 4.9   &18 10 28.27 &$-$19 22 31.1 &  27.0(5.8)                 	 & $<$0.2	 \\[3pt]  
	         & 7.4   &18 10 28.26 & $-$19 22 31.2 & 21.7(6.3)          		& 			 \\[3pt]  
	         & 20.9 & \nodata	& \nodata		  & $<33\tablenotemark{a}  $             		&        		  \\[3pt]  
	         & 25.5 & \nodata	 & \nodata	           & $<36\tablenotemark{a}  $             		&        		  \\[3pt]

\enddata

\tablenotetext{a}{Non-detection. Upper limit is 3$\sigma$ value in map.}
\tablenotetext{b}{Reported fluxes for component C contain a contribution from the extended weak component connecting components B and C.}
\tablecomments{1$\sigma$ uncertainties are reported.}

\end{deluxetable*}

\normalsize

% Reset the footnotes back to numbers
\renewcommand{\thefootnote}{\arabic{footnote}}

%%%%%%%%%%%%%%%%%%%%%%%%%%%%%%%%%%%%%%%%%%%%%

%%%%%%%%%%%%%%%

%%%%%%%%%%%%%%%

\section{Analysis}\label{analysis}

The above results clearly show that a string of radio sources is associated with the massive proto-stellar
object in  G11P1. In this section we will discuss the physical nature of the emission, and present archival infrared data.

\subsection{Radio Continuum}\label{rad_cont}
Components B and C appear connected at $1.3\,$cm, however this bridge of emission is not detected
at $6\,$cm. For our analysis we will not consider this bridging structure, thus component B and C are treated as individual sources.
First, we might consider that radio components A to D are manifestations of individual massive stars which ionize
their surroundings, i.e. ultra or hyper compact HII regions (UCHII or HCHII). Due to the much improved continuum sensitivity of the VLA, it should be now
possible to explore photo-ionized regions around stars of spectral type later than B2 throughout the Galaxy. 
The orientation of the 4 putative stars is approximately along the dark filament
as might be expected for star formation in this environment. A similar alignment of proto-stellar objects along the dark cloud
has for instance been observed in the region G28.34+0.06 (e.g. \citealt{2011ApJ...735...64W}). However, in the latter case the sources have separations
of the order of $0.1\,$pc, whereas in G11P1 there are 4 sources within a distance of $9000\,$AU (0.04 pc at the distance of G11P1). 
Several massive objects can in fact be found at such small and even smaller separation in young clusters (e.g. Orion Trapezium; NGC$\,$2071: \citealt{2012ApJ...746...71C}).
We will thus first consider the formation of 4 massive stars aligned along the filament. 

An argument against the hypothesis that the 4 radio sources are ionized by 4 individual stars can be made 
if we consider the implied luminosities from 4 individual stars. Assuming optically thin free-free emission and neglecting absorption of ionizing
photons within the UC/HCHII regions we calculated the Lyman continuum luminosity for each component using the formulas given in \citet{1994ApJS...91..659K}.
Using the tabulation in \citet{2005IAUS..227..389C}, the corresponding spectral types are approximately B2/B3 for each radio component. Due to the
assumptions made in the calculation, these values are lower limits. Such stars have a luminosity of $> 1000\,$L$_{\odot}$, hence for four such
stars we would predict a luminosity of the region of more than  $4000\,$L$_{\odot}$, which is much larger than the measured luminosity of the region of about
1200 L$_{\odot}$ (P06) or 1346 L$_{\odot}$ \citep{2010A&A...518L..95H}. Therefore the hypothesis of 4 individual UC/HCHII regions can be excluded. 

\begin{figure}
\centering

\includegraphics[scale=0.5]{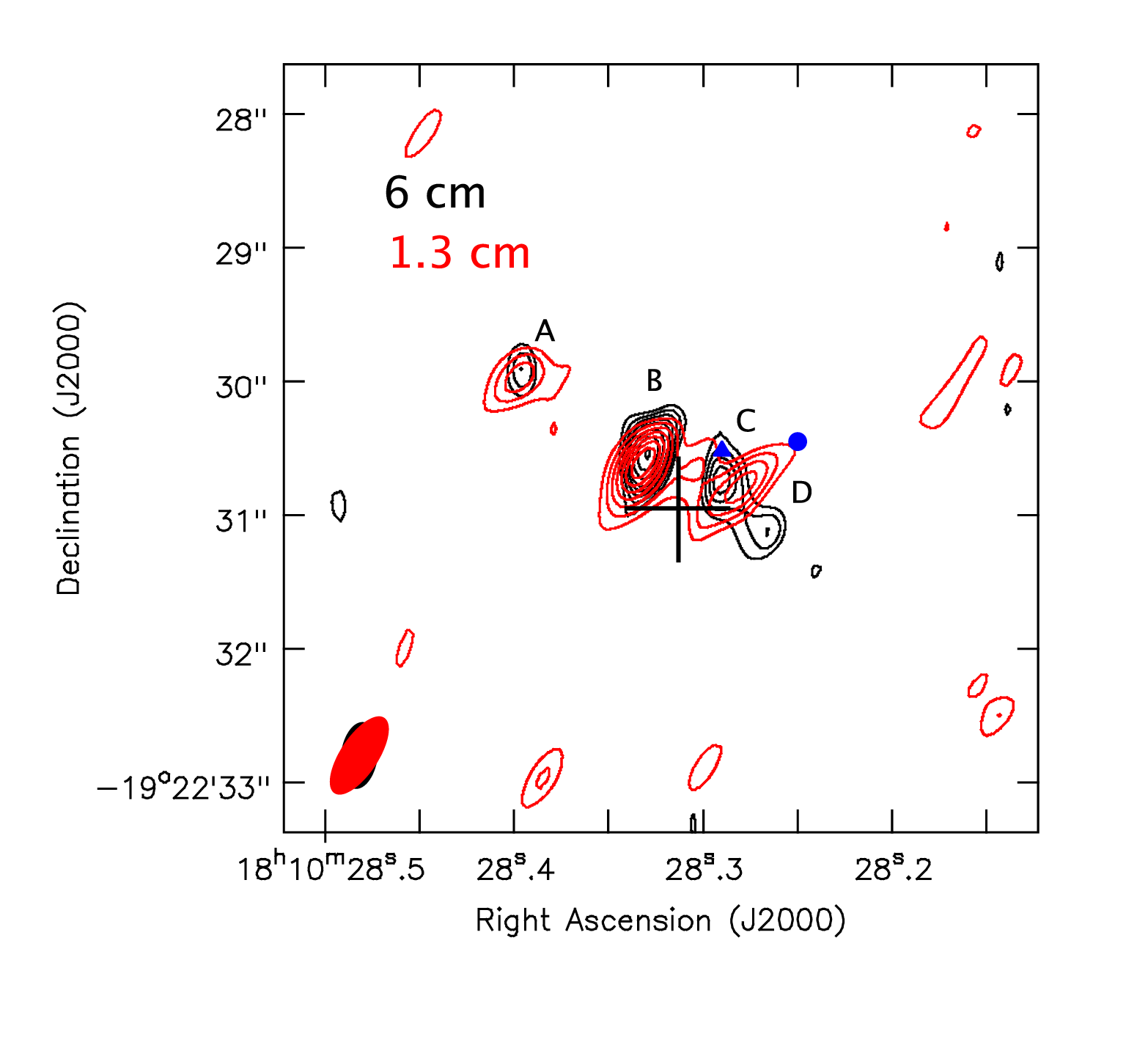}

\caption{6 cm (black) and 1.3 cm (red)  contours of the VLA observations toward
G11.11$-$0.12P1.  The blue filled circle indicates the CH$_{3}$OH maser at 
RA(J2000)=$18^{h}10^{m}28{\rlap.}^{s}25$, Dec(J2000)$=-$$19^{\circ}22^{\prime}30{\rlap.}^{\prime\prime}45 $
and the blue filled triangle indicates the H$_{2}$O maser at 
RA(J2000)=$18^{h}10^{m}28{\rlap.}^{s}29$, Dec(J2000)$=-$$19^{\circ}22^{\prime}30{\rlap.}^{\prime\prime}5 $,
detected by P06. The black cross corresponds to 
the position of the IR source SSTGLMC G011.1089-00.1144 detected at 3.6 $\mu$m from the IRAC GLIMPSE images (Figure~\ref{f1})}.\label{f4}
\end{figure}

Next, we can ask whether external photoionization can explain the 4 radio sources. In this scenario 4 clumps are externally ionized by an unseen massive
protostar. The position of the putative accretion disk traced by $6.7\,$GHz methanol masers (P06) lies somewhat offset from the line
defined by the 4 radio sources (see Figure \ref{f4}). We calculated the necessary flux of ionizing photons correcting for the
ratio of solid angle $\Omega/4\pi$ of the radio sources as seen from the position of the methanol maser source. For the calculation we 
assumed source sizes of half of the synthesized beam, likely an overestimate resulting in a lower limit on the corrected ionizing flux.
To externally ionize the four components we find that a single star of spectral type B1 or earlier would be required. Such stars have luminosities
of $>5000$ L$_{\odot}$, which is also in conflict with the measured luminosity of the region.  A calculation placing the star at the peak position of
radio continuum source B gives similar results. We conclude that direct photoionization of the cm components from a single massive proto-stellar objects is unlikely.

We suggest therefore that the radio continuum emission detected towards G11P1 is produced by shock ionization. 
This could be the result of either accretion shocks caused by supersonic infall onto an accretion disk, or shocks caused by the interaction of a stellar wind 
with surrounding molecular core matter. The expected radio continuum emission from accretion shocks has been calculated by \citet{1996ApJ...471L..45N}; at the
distance of G11P1, and assuming a mass of $8\,$M$_\odot$ (P06), their model predicts a $4.8\,$GHz flux density of below $1\,\mu$Jy for an accretion rate
of $10^{-4}\,$M$_\odot\,$yr$^{-1}$. Thus, unless one wants to accept an unusually large accretion rate, the accretion shock scenario seems to be ruled out. 
On the other hand, a scenario where a neutral wind driven by the 
embedded massive protostar shocks against surrounding  high-density matter and produces free-free emission (\citealt{1987RMxAA..14..595C} and references therein),
appears more likely. 

Before discussing this scenario in more detail below, we note that the above luminosity argument could also be made consistent with the data if we assume that only one of the
four radio sources is  a UC/HCHII region and the
other sources are shock-ionized. The spectral behavior of component B is close to that of an optically thin HII region, hence we have considered this possibility as well.
P06 and \citet{2011A&A...529A.161G} have estimated a molecular hydrogen density of $7\times 10^5\,$cm$^{-3}$, and a temperature of $60\,$K for the G11P1 central core.
Including also the turbulent pressure of the molecular gas given by the FWHM of the hot NH$_3$ component ($4\,$km$\,$s$^{-1}$), we have calculated
the size of an UC/HCHII region around a B3 ZAMS star 
given by the condition of pressure equilibrium between molecular and ionized gas using the formulas of \citet{1996ApJ...473L.131X}. We obtain a size
of about $400\,$AU, which is consistent with the region being unresolved in our observations. Thus, our data do not exclude a UC/HCHII region interpretation
for component B (only), plus cm emission from 3 shocked regions. These 3 continuum sources could then be either separate jets from individual proto-stars which are unresolved,
or several shocks from a single jet, likely caused by episodic matter ejection. 
We also note that if one adopts the empirical correlation between radio and bolometric luminosity of \citet{2007ApJ...667..329S} an interpretation of the
radio sources as four independent lower mass stars is not excluded by the measured luminosity.

Because of the alignment and orientation of the 4 radio sources with respect to several disk tracers (see below) we will 
favor shock ionization from a single jet as the likely physical scenario for the cm emission. In this picture a
massive star in or near component B drives a bipolar jet which causes the observed radio emission when the ejecta interact with the surrounding core matter.   
Assuming then that the radio emission detected at G11P1 originates from a jet, we use the standard model of  \citet{1986ApJ...304..713R} for free-free emission 
of a collimated, ionized flow or wind, with constant velocity, temperature and 
ionization fraction. Reynolds' model suggests that the observed flux density and the angular size depend on frequency as $S_{\nu} \propto \nu^{1.3-0.7/\epsilon}$ and 
$\theta_{maj} \propto \nu^{-0.7/\epsilon}$, where $\epsilon$ depends on the geometry of the jet and is the power-law index that describes 
the dependence of the jet half-width on the distance from the jet origin \citep{1986ApJ...304..713R}. For the G11P1 component B, the observed dependence of 
the flux density with frequency gives a value of $\epsilon \sim0.6$, which within the uncertainties, is in agreement with a  collimated 
ionized jet. The angular size dependence with frequency cannot be determined for any of the components since they are all unresolved with our angular resolution. 
Using equation 19 from  \citet{1986ApJ...304..713R} we can make a rough  estimate of the mass-loss rate ($\dot{M}$) of the G11P1 B-component,
assuming parameter values that are typical of jets associated with luminous objects ($v_{wind}$= 700 km s$^{-1}$; $\theta_{0}=1\, $rad; T$_{e}=10^{4}$ K, $i=45^{\circ}$,  $x_{0}=0.1$; e.g. \citealt{1994ApJ...430L..65R}).
The estimated mass loss rate 
of G11P1 component B observed at 25.5 GHz is $\dot{M} \sim 3 \times 10^{-6}$  M$_{\odot}$\,yr$^{-1}$. We can get an estimated value of the momentum rate ($\dot{P}$) 
by multiplying the mass loss rate by the typical velocity of the wind in massive stars, which gives  $\dot{P} \sim 2
 \times 10^{-3}$ M$_{\odot}$\,yr$^{-1}$ km\,s$^{-1}$. On the other hand, if we assume that component B is a jet produced by shock induced ionization \citep[e.g.][]{1987RMxAA..14..595C, 2013A&A...551A..43J} with a shock efficiency ($\eta$) $\sim$0.1 and an optical depth of the emission at 25.5 GHz of ~0.02, the estimated momentum rate is $\dot{P} \sim 7 \times 10^{-3}$ M$_{\odot}$\,yr$^{-1}$ km\,s$^{-1}$. The fact that  $\dot{P}$ estimated from the shock ionization mechanism is $\sim4$ times bigger than the one predicted by Reynolds' model  suggest  that the 25.5 GHz emission is not completely due to shocks in the jet. However, the momentum rate of the jet could ionize itself  if the jet velocity is $\sim$1600 km\,s$^{-1}$ or  the shock efficiency is $\sim$0.4.

%%%%%%%%%%%%%%%%%%%%%%%%%%

%%%%%%%%%%%%%%%%%%%%%%%%%%

\subsection{Near-IR Sources}\label{near_ir}
P06 proposed from their study of the spectral energy distribution of  G11P1
the presence of an accretion disk around the central massive young star. From 2MASS data (resolution $\sim$2$^{\prime \prime}$),  P06 detected three faint sources in the J and 
H bands with only  one of those sources  detected  at K-band. In their analysis they found that these sources cannot be explained by reddening and they proposed 
that those NIR detections are knots of scattered light that escape from the star into an optically thin cone
above a circumstellar disk.

We retrieved data from the UKIRT Infrared Deep Sky Survey (UKIDSS) GPS   
and compared them with the corresponding 2MASS data analyzed and discussed by P06. 
The UKIDSS project is defined in \citet{2007MNRAS.379.1599L}. UKIDSS uses the UKIRT
Wide Field Camera (WFCAM; \citet{2007A&A...467..777C}). The photometric system is
described in \citet{2006MNRAS.367..454H}, and the calibration is described in \citet{2009MNRAS.394..675H}. The pipeline processing and science archive are described in
\citet{2009Icar..203..287I} and \citet{2008MNRAS.384..637H}.
The UKIDSS data are three magnitudes deeper and have higher angular resolution ($\sim0.4 ^{\prime \prime}$)
compared to 2MASS data. The astrometric accuracy of the UKIDSS data is about 50 mas.

\begin{figure*}
\centering

\includegraphics[scale=0.25]{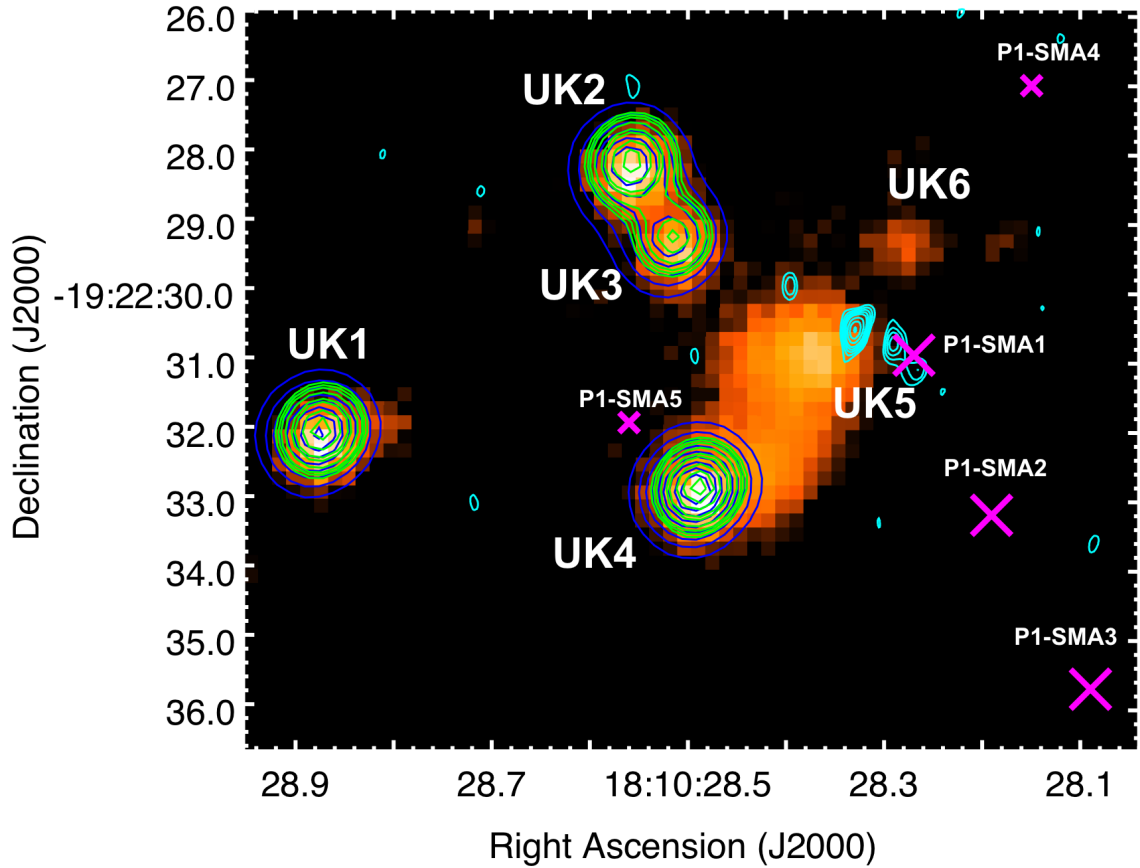}

\caption{UKIDSS K-band image of the G11.11$-0.121\,$P1 region. Overlaid are in contours UKIDSS J-band (blue) and H-band (green) emission. Six sources are detected in total, where two of them are only detected in the  K-band. Sources are labeled with the UK prefix. VLA 6 cm continuum emission contours (cyan) are overlaid. Magenta crosses represent the SMA condensations detected by \citet{2014MNRAS.439.3275W}. Condensation P1-SMA6 is not shown in this figure.}\label{f5}
\end{figure*}

We found a total of six  sources associated with the G11P1 core, two of which are only seen at K-band (see Figure \ref{f5}). 
The sources are labeled as UK1 to UK6.
Based on their positions in a JHK color-color diagram  we found that UK1 and UK2 can be explained as main sequence stars  with a visual extinction $\le 7$ mag. Thus, we suggest that these two components are foreground stars. 
Not much can be said about UK3 since its H-band magnitude is not available. 
On the other hand, the position of UK4 in the JHK color-color diagram indicates intrinsic IR excess emission and therefore we suggest that UK4 is likely a young star. Thus, 
the 2MASS sources from P06 associated with UK1, UK2 and UK4 do not appear to be knots of scattered light, but appear to be of stellar nature. On the other hand, components UK5 and UK6 are only detected at K-band  which indicates very high extinction. UK5 is an extended source oriented in the direction toward UK6, along
a PA of $\sim 130^{\circ}$. These two sources are separated by an angular distance
of $\sim 2.2^{\prime \prime}$ ($\sim$7900 AU at the distance of 3.6 kpc).
Figure \ref{f5} shows that these 2 components are oriented roughly perpendicular to the axis defined by the radio continuum components 
which we believe is caused by a bipolar jet. We also note that the Mid-IR {\it Spitzer} IRAC data are clearly offset from the NIR sources and peak closer to the radio data.
Considering the excellent astrometrical agreement between {\it Spitzer} IRAC and UKIDSS data, we are convinced that the offset is real, which would speak against a YSO nature of
UK5 and UK6, but is consistent with their K-band emission coming from scattered light from an accretion disk. 
Therefore, while different in detail, the higher quality UKIDSS data are
supportive of the interpretation of P06 for the presence of a disk-like structure in the G11P1 core.\\
Another interpretation for UK5 and UK6 is that they are scattered light at the inner wall cavity produced by the molecular outflow. Figure \ref{f5} shows that the outflow cavity appears brighter to the east which is consistent with the blueshifted SiO outflow from Wang et al. 2014. In this picture, the redshifted side at NIR (i.e. UK6) is fainter due to dependence of cavity brightness with inclination \citep[e.g.][]{2008ApJ...679.1364T}.

%%%%%%%%%%%%%%%%%%%%%%%%%%%%%%%%%

%%%%%%%%%%%%%%%%%%%%%%%%%%%%%%%%%

\section {Discussion}\label{discussion}

Several authors (e.g. P06, \citealt{2011A&A...529A.161G}) have interpreted the G11P1 central object as a massive protostar in a very early stage of evolution. The detection
of a mid-IR point-source and the measured luminosity of around $1000\,$L$_\odot$ clearly indicate the presence of a stellar object, whose energy output is comparable
to an $8\,$M$_\odot$ ZAMS star (P06). The presence of the $6.7\,$GHz CH$_{3}$OH  and $22\,$GHz H$_{2}$O masers are strong indicators of massive star formation, and the presence
of a massive core (e.g. 240 M$_{\odot}$, \citealt{2010A&A...518L..95H}) allows in principle to accrete more mass. The relative youth of this system is demonstrated by the fact
that most of the  molecular gas in the core appears to be quite cold: P06 report a temperature of only $15.4\,$K for the overall molecular core based on NH$_{3}$ observations.
Whether or not the central object in G11P1 is in fact accumulating more mass and will grow to a massive star can in principle be observationally decided by the detection
of outflow activity, because flows and jets are thought to be intimately linked to mass accretion.  The molecular line observations of \citealt{2011A&A...529A.161G} resulted in the detection of 
non-gaussian line wings, and a possible outflow traced by the CH$_{3}$OH(2$_{k} - 1_{k}$) lines. This was recently confirmed by \citet{2014MNRAS.439.3275W} who found an East--West outflow
in the SiO(5--4) line. In the previous section we have argued that the radio continuum emission
from G11P1 is best explained by an ionized jet, and we hence add to the picture an outflow tracer very near the protostar.

How do the results described in this paper fit into the picture of a massive protostar with a disk/jet system as defined by previous observations?
We first note that the $6.7\,$GHz CH$_{3}$OH maser is located
not on the axis defined by the radio continuum sources (see Fig.~\ref{f4}), but is offset by about $0.8^{\prime\prime}$. According to P06 the astrometrical uncertainty of their
measurement is of that order, so that the maser could in fact be located nearer to the jet axis as expected if the maser spots arise in an accretion disk. 
Fitting individual gaussians
to different channels within the maser line P06 find a linear structure of length $0.2^{\prime\prime}$ oriented approximately North--South. Such linear structures have
often been observed for the $6.7\,$GHz CH$_{3}$OH maser line (e.g. \citet{2000A&A...362.1093M}), however the interpretation as disk tracer is not unique, as similar
structures are expected in shock fronts associated with outflows. In fact, in the scenario where the NIR knots (UK5 and UK6) are scattered light at the outflow cavities, a linearly distributed maser emission might be then tracing the walls of the outflow cavities \citep[e.g.][]{2005ApJ...628L.151D, 2011MNRAS.410..627T}. 
We also note that there is a maser listed in the $6.7\,$GHz methanol multi-beam maser catalogue of \citet{2010MNRAS.409..913G} which position is different by more than $1^{\prime\prime}$ from
the position given in P06. To clarify these issues, higher angular resolution observations of this maser with high astrometric precision are needed. 

As mentioned above, the CH$_{3}$OH and SiO line observations of \citet{2011A&A...529A.161G} and \citet{2014MNRAS.439.3275W}, respectively, suggest a molecular flow from the massive star in G11P1
along the East--West direction, and this
orientation is consistent with the North-South orientation of the CH$_3$OH maser disk postulated by P06. The outflow direction defined by the cm continuum sources is closer
to NE-SW, and in this case we also have a perpendicular component tracing a possible disk, namely the NIR emission discussed in the previous section. Additional evidence for
this disk component comes from the recent SMA and VLA observations of \citet{2014MNRAS.439.3275W}. Their Figure~11 shows that the $880\,\mu$m dust emission is oriented  nearly 
perpendicular to the ionized jet, and the NH$_3$(2,2) line shows a clear velocity gradient along that same direction. Both dust emission and NH$_3$(2,2) are more likely to trace
dense matter present in a disk/torus system than a molecular outflow.

This apparent contradiction of the different flow orientations can then be explained in at least two ways. First, it is possible that a second protostar is responsible
for the East-West outflow, whereas the flow associated with the radio jet has not been detected.
Second, a change of alignment of flow axis on different length scales is a known phenomenon, e.g. in the case
of the massive protostar in IRAS$\,$20126$+$4104, the jet axis changes from a NW--SE orientation on arc-second scales to a North--South direction when probed by CO on
arc-minutes scales \citep{2000ApJ...535..833S}. Another well known case where a misalignment of outflow axis is observed is the protostar NGC$\,$7538 IRS1;
\citet{2006A&A...455..521K} describe a number of possible disk precession mechanisms which could explain such changes in the flow axis. If this is the the case for the G11P1
protostar, the outflow angle would have to change by $\sim50^{\circ}$ from the $1^{\prime\prime}$ scale of the ionized jet, to the $10^{\prime\prime}$ scale where the molecular flows
have been detected in SiO and CH$_3$OH.

Previous studies have shown that all high-mass YSOs associated with ionized jets are also associated with large scale, high velocity collimated molecular outflows,
and that there exists a correlation between the radio luminosity, and the momentum rate of the molecular outflows (e.g. \citealt{1992ApJ...395..494A}). 
The $4.9\,$GHz radio luminosity of G11P1 is relatively large with
S$_{\nu}$d$^{2}=2.6$ mJy$\,$kpc$^{2}$, near the lower range of what is observed for jets from massive protostars \citep{2008AJ....135.2370R},
but much larger than radio luminosities from low mass stars. 
In section~\ref{rad_cont} we have estimated mass loss
and momentum rates for the jet in G11P1. The resulting values are lower than what is found for jets from massive protostars like
IRAS$\,$16547$-$4247 \citep{2003ApJ...587..739G}, or IRAS$\,$16562$-$3959 \citep{2010ApJ...725..734G}, but there are still many uncertain
assumptions made in the estimate of these quantities. Furthermore, these sources are much more luminous than G11P1, and we can speculate
that the lower values for mass loss and momentum rates are due to the earlier evolutionary state of the protostar in G11P1. 
If we assume
that the molecular flow observed by \citet{2014MNRAS.439.3275W} is related to the radio luminosity, we find that the jet/flow data of G11P1 fall close to
the radio luminosity/momentum rate relation of  \cite{1992ApJ...395..494A}.

\section {Conclusion}\label{conclusions}

Previous observations have established the stellar source in the G11P1 core as a new candidate for a massive 
protostar in a very early evolutionary stage. The upgraded VLA provides the high sensitivity to detect these types of very early 
massive protostars in the radio continuum down to a rms of few $\mu$Jy/beam.
Our VLA continuum observations
reveal four weak, and unresolved sources, centered on the mid-IR source, which are aligned in a NE-SW direction.
The spectral indices determined for each component are consistent with partially optically thick ($\alpha > -0.1$) free-free emission from ionized
gas arising from a thermal jet \citep{1998AJ....116.2953A}, where the mechanism of ionization is most likely by shock ionization \citep{1987RMxAA..14..595C,1989ApL&C..27..299C}.
We also present  archival NIR data from UKIRT (resolution of $\sim 0.4^{\prime \prime}$). These data reveal an extended structure only visible in K-band, 
which is oriented perpendicular to the orientation of the radio continuum data. This structure can be interpreted as scattered light from an
accretion disk. Our observations thus provide new evidence that a disk/jet system is present in the protostar in G11P1.

\acknowledgements

We thank T. Pillai and L. G\'omez for  helpful discussions. PH acknowledges partial support from NSF grant AST-0908901. We thank J. Marvil, U. Rau and E. Momjian at NRAO, Socorro for stimulating technical discussions about the VLA capabilities.
Some of the data reported here were obtained as part of the UKIRT Service Program. The United Kingdom Infrared Telescope is operated by the Joint Astronomy Centre on behalf of the UK Particle Physics and Astronomy Research Council.  We thank the anonymous referee whose suggestions improved this manuscript.

%\clearpage 


\begin{thebibliography}{43}
\expandafter\ifx\csname natexlab\endcsname\relax\def\natexlab#1{#1}\fi

\bibitem[{{Anglada} {et~al.}(1992){Anglada}, {Rodriguez}, {Canto}, {Estalella},
  \& {Torrelles}}]{1992ApJ...395..494A}
{Anglada}, G., {Rodriguez}, L.~F., {Canto}, J., {Estalella}, R., \&
  {Torrelles}, J.~M. 1992, \apj, 395, 494

\bibitem[{{Anglada} {et~al.}(1998){Anglada}, {Villuendas}, {Estalella},
  {Beltr{\'a}n}, {Rodr{\'{\i}}guez}, {Torrelles}, \&
  {Curiel}}]{1998AJ....116.2953A}
{Anglada}, G., {Villuendas}, E., {Estalella}, R., {et~al.} 1998, \aj, 116, 2953

\bibitem[{{Breen} {et~al.}(2013){Breen}, {Ellingsen}, {Contreras}, {Green},
  {Caswell}, {Stevens}, {Dawson}, \& {Voronkov}}]{2013MNRAS.435..524B}
{Breen}, S.~L., {Ellingsen}, S.~P., {Contreras}, Y., {et~al.} 2013, \mnras,
  435, 524

\bibitem[{{Butler} {et~al.}(2014){Butler}, {Tan}, \&
  {Kainulainen}}]{2014ApJ...782L..30B}
{Butler}, M.~J., {Tan}, J.~C., \& {Kainulainen}, J. 2014, \apjl, 782, L30

\bibitem[{{Carey} {et~al.}(1998){Carey}, {Clark}, {Egan}, {Price}, {Shipman},
  \& {Kuchar}}]{1998ApJ...508..721C}
{Carey}, S.~J., {Clark}, F.~O., {Egan}, M.~P., {et~al.} 1998, \apj, 508, 721

\bibitem[{{Carey} {et~al.}(2000){Carey}, {Feldman}, {Redman}, {Egan},
  {MacLeod}, \& {Price}}]{2000ApJ...543L.157C}
{Carey}, S.~J., {Feldman}, P.~A., {Redman}, R.~O., {et~al.} 2000, \apjl, 543,
  L157

\bibitem[{{Carrasco-Gonz{\'a}lez} {et~al.}(2012){Carrasco-Gonz{\'a}lez},
  {Osorio}, {Anglada}, {D'Alessio}, {Rodr{\'{\i}}guez}, {G{\'o}mez}, \&
  {Torrelles}}]{2012ApJ...746...71C}
{Carrasco-Gonz{\'a}lez}, C., {Osorio}, M., {Anglada}, G., {et~al.} 2012, \apj,
  746, 71

\bibitem[{{Casali} {et~al.}(2007){Casali}, {Adamson}, {Alves de Oliveira},
  {Almaini}, {Burch}, {Chuter}, {Elliot}, {Folger}, {Foucaud}, {Hambly},
  {Hastie}, {Henry}, {Hirst}, {Irwin}, {Ives}, {Lawrence}, {Laidlaw}, {Lee},
  {Lewis}, {Lunney}, {McLay}, {Montgomery}, {Pickup}, {Read}, {Rees}, {Robson},
  {Sekiguchi}, {Vick}, {Warren}, \& {Woodward}}]{2007A&A...467..777C}
{Casali}, M., {Adamson}, A., {Alves de Oliveira}, C., {et~al.} 2007, \aap, 467,
  777

\bibitem[{{Crowther}(2005)}]{2005IAUS..227..389C}
{Crowther}, P.~A. 2005, in IAU Symposium, Vol. 227, Massive Star Birth: A
  Crossroads of Astrophysics, ed. R.~{Cesaroni}, M.~{Felli}, E.~{Churchwell},
  \& M.~{Walmsley}, 389--396

\bibitem[{{Curiel} {et~al.}(1987){Curiel}, {Canto}, \&
  {Rodriguez}}]{1987RMxAA..14..595C}
{Curiel}, S., {Canto}, J., \& {Rodriguez}, L.~F. 1987, \RMxAA, 14, 595

\bibitem[{{Curiel} {et~al.}(1989){Curiel}, {Rodriguez}, {Bohigas}, {Roth},
  {Canto}, \& {Torrelles}}]{1989ApL&C..27..299C}
{Curiel}, S., {Rodriguez}, L.~F., {Bohigas}, J., {et~al.} 1989, Astrophysical
  Letters and Communications, 27, 299

\bibitem[{{Cyganowski} {et~al.}(2008){Cyganowski}, {Whitney}, {Holden},
  {Braden}, {Brogan}, {Churchwell}, {Indebetouw}, {Watson}, {Babler},
  {Benjamin}, {Gomez}, {Meade}, {Povich}, {Robitaille}, \&
  {Watson}}]{2008AJ....136.2391C}
{Cyganowski}, C.~J., {Whitney}, B.~A., {Holden}, E., {et~al.} 2008, \aj, 136,
  2391

\bibitem[{{De Buizer} \& {Minier}(2005)}]{2005ApJ...628L.151D}
{De Buizer}, J.~M., \& {Minier}, V. 2005, \apjl, 628, L151

\bibitem[{{Garay} {et~al.}(2003){Garay}, {Brooks}, {Mardones}, \&
  {Norris}}]{2003ApJ...587..739G}
{Garay}, G., {Brooks}, K.~J., {Mardones}, D., \& {Norris}, R.~P. 2003, \apj,
  587, 739

\bibitem[{{G{\'o}mez} {et~al.}(2011){G{\'o}mez}, {Wyrowski}, {Pillai},
  {Leurini}, \& {Menten}}]{2011A&A...529A.161G}
{G{\'o}mez}, L., {Wyrowski}, F., {Pillai}, T., {Leurini}, S., \& {Menten},
  K.~M. 2011, \aap, 529, A161

\bibitem[{{Green} {et~al.}(2010){Green}, {Caswell}, {Fuller}, {Avison},
  {Breen}, {Ellingsen}, {Gray}, {Pestalozzi}, {Quinn}, {Thompson}, \&
  {Voronkov}}]{2010MNRAS.409..913G}
{Green}, J.~A., {Caswell}, J.~L., {Fuller}, G.~A., {et~al.} 2010, \mnras, 409,
  913

\bibitem[{{Guzm{\'a}n} {et~al.}(2010){Guzm{\'a}n}, {Garay}, \&
  {Brooks}}]{2010ApJ...725..734G}
{Guzm{\'a}n}, A.~E., {Garay}, G., \& {Brooks}, K.~J. 2010, \apj, 725, 734

\bibitem[{{Hambly} {et~al.}(2008){Hambly}, {Collins}, {Cross}, {Mann}, {Read},
  {Sutorius}, {Bond}, {Bryant}, {Emerson}, {Lawrence}, {Rimoldini}, {Stewart},
  {Williams}, {Adamson}, {Hirst}, {Dye}, \& {Warren}}]{2008MNRAS.384..637H}
{Hambly}, N.~C., {Collins}, R.~S., {Cross}, N.~J.~G., {et~al.} 2008, \mnras,
  384, 637

\bibitem[{{Henning} {et~al.}(2010){Henning}, {Linz}, {Krause}, {Ragan},
  {Beuther}, {Launhardt}, {Nielbock}, \& {Vasyunina}}]{2010A&A...518L..95H}
{Henning}, T., {Linz}, H., {Krause}, O., {et~al.} 2010, \aap, 518, L95

\bibitem[{{Hewett} {et~al.}(2006){Hewett}, {Warren}, {Leggett}, \&
  {Hodgkin}}]{2006MNRAS.367..454H}
{Hewett}, P.~C., {Warren}, S.~J., {Leggett}, S.~K., \& {Hodgkin}, S.~T. 2006,
  \mnras, 367, 454

\bibitem[{{Hodgkin} {et~al.}(2009){Hodgkin}, {Irwin}, {Hewett}, \&
  {Warren}}]{2009MNRAS.394..675H}
{Hodgkin}, S.~T., {Irwin}, M.~J., {Hewett}, P.~C., \& {Warren}, S.~J. 2009,
  \mnras, 394, 675

\bibitem[{{Irwin} {et~al.}(2009){Irwin}, {Teanby}, \&
  {Davis}}]{2009Icar..203..287I}
{Irwin}, P.~G.~J., {Teanby}, N.~A., \& {Davis}, G.~R. 2009, \icarus, 203, 287

\bibitem[{{Johnston} {et~al.}(2013){Johnston}, {Shepherd}, {Robitaille}, \&
  {Wood}}]{2013A&A...551A..43J}
{Johnston}, K.~G., {Shepherd}, D.~S., {Robitaille}, T.~P., \& {Wood}, K. 2013,
  \aap, 551, A43

\bibitem[{{Johnstone} {et~al.}(2003){Johnstone}, {Fiege}, {Redman}, {Feldman},
  \& {Carey}}]{2003ApJ...588L..37J}
{Johnstone}, D., {Fiege}, J.~D., {Redman}, R.~O., {Feldman}, P.~A., \& {Carey},
  S.~J. 2003, \apjl, 588, L37

\bibitem[{{Kraus} {et~al.}(2006){Kraus}, {Balega}, {Elitzur}, {Hofmann},
  {Preibisch}, {Rosen}, {Schertl}, {Weigelt}, \& {Young}}]{2006A&A...455..521K}
{Kraus}, S., {Balega}, Y., {Elitzur}, M., {et~al.} 2006, \aap, 455, 521

\bibitem[{{Kurtz} {et~al.}(1994){Kurtz}, {Churchwell}, \&
  {Wood}}]{1994ApJS...91..659K}
{Kurtz}, S., {Churchwell}, E., \& {Wood}, D.~O.~S. 1994, \apjs, 91, 659

\bibitem[{{Lawrence} {et~al.}(2007){Lawrence}, {Warren}, {Almaini}, {Edge},
  {Hambly}, {Jameson}, {Lucas}, {Casali}, {Adamson}, {Dye}, {Emerson},
  {Foucaud}, {Hewett}, {Hirst}, {Hodgkin}, {Irwin}, {Lodieu}, {McMahon},
  {Simpson}, {Smail}, {Mortlock}, \& {Folger}}]{2007MNRAS.379.1599L}
{Lawrence}, A., {Warren}, S.~J., {Almaini}, O., {et~al.} 2007, \mnras, 379,
  1599

\bibitem[{{Menten} {et~al.}(2005){Menten}, {Pillai}, \&
  {Wyrowski}}]{2005IAUS..227...23M}
{Menten}, K.~M., {Pillai}, T., \& {Wyrowski}, F. 2005, in IAU Symposium, Vol.
  227, Massive Star Birth: A Crossroads of Astrophysics, ed. R.~{Cesaroni},
  M.~{Felli}, E.~{Churchwell}, \& M.~{Walmsley}, 23--34

\bibitem[{{Minier} {et~al.}(2000){Minier}, {Booth}, \&
  {Conway}}]{2000A&A...362.1093M}
{Minier}, V., {Booth}, R.~S., \& {Conway}, J.~E. 2000, \aap, 362, 1093

\bibitem[{{Minier} {et~al.}(2003){Minier}, {Ellingsen}, {Norris}, \&
  {Booth}}]{2003A&A...403.1095M}
{Minier}, V., {Ellingsen}, S.~P., {Norris}, R.~P., \& {Booth}, R.~S. 2003,
  \aap, 403, 1095

\bibitem[{{Neufeld} \& {Hollenbach}(1996)}]{1996ApJ...471L..45N}
{Neufeld}, D.~A., \& {Hollenbach}, D.~J. 1996, \apjl, 471, L45

\bibitem[{{Pillai} {et~al.}(2006){Pillai}, {Wyrowski}, {Menten}, \&
  {Kr{\"u}gel}}]{2006A&A...447..929P}
{Pillai}, T., {Wyrowski}, F., {Menten}, K.~M., \& {Kr{\"u}gel}, E. 2006, \aap,
  447, 929

\bibitem[{{Rathborne} {et~al.}(2006){Rathborne}, {Jackson}, \&
  {Simon}}]{2006ApJ...641..389R}
{Rathborne}, J.~M., {Jackson}, J.~M., \& {Simon}, R. 2006, \apj, 641, 389

\bibitem[{{Reynolds}(1986)}]{1986ApJ...304..713R}
{Reynolds}, S.~P. 1986, \apj, 304, 713

\bibitem[{{Rodriguez} {et~al.}(1994){Rodriguez}, {Garay}, {Curiel}, {Ramirez},
  {Torrelles}, {Gomez}, \& {Velazquez}}]{1994ApJ...430L..65R}
{Rodriguez}, L.~F., {Garay}, G., {Curiel}, S., {et~al.} 1994, \apjl, 430, L65

\bibitem[{{Rodr{\'{\i}}guez} {et~al.}(2008){Rodr{\'{\i}}guez}, {Moran},
  {Franco-Hern{\'a}ndez}, {Garay}, {Brooks}, \&
  {Mardones}}]{2008AJ....135.2370R}
{Rodr{\'{\i}}guez}, L.~F., {Moran}, J.~M., {Franco-Hern{\'a}ndez}, R., {et~al.}
  2008, \aj, 135, 2370

\bibitem[{{Shepherd} {et~al.}(2000){Shepherd}, {Yu}, {Bally}, \&
  {Testi}}]{2000ApJ...535..833S}
{Shepherd}, D.~S., {Yu}, K.~C., {Bally}, J., \& {Testi}, L. 2000, \apj, 535,
  833

\bibitem[{{Shirley} {et~al.}(2007){Shirley}, {Claussen}, {Bourke}, {Young}, \&
  {Blake}}]{2007ApJ...667..329S}
{Shirley}, Y.~L., {Claussen}, M.~J., {Bourke}, T.~L., {Young}, C.~H., \&
  {Blake}, G.~A. 2007, \apj, 667, 329

\bibitem[{{Tobin} {et~al.}(2008){Tobin}, {Hartmann}, {Calvet}, \&
  {D'Alessio}}]{2008ApJ...679.1364T}
{Tobin}, J.~J., {Hartmann}, L., {Calvet}, N., \& {D'Alessio}, P. 2008, \apj,
  679, 1364

\bibitem[{{Torrelles} {et~al.}(2011){Torrelles}, {Patel}, {Curiel},
  {Estalella}, {G{\'o}mez}, {Rodr{\'{\i}}guez}, {Cant{\'o}}, {Anglada},
  {Vlemmings}, {Garay}, {Raga}, \& {Ho}}]{2011MNRAS.410..627T}
{Torrelles}, J.~M., {Patel}, N.~A., {Curiel}, S., {et~al.} 2011, \mnras, 410,
  627

\bibitem[{{Wang} {et~al.}(2011){Wang}, {Zhang}, {Wu}, \&
  {Zhang}}]{2011ApJ...735...64W}
{Wang}, K., {Zhang}, Q., {Wu}, Y., \& {Zhang}, H. 2011, \apj, 735, 64

\bibitem[{{Wang} {et~al.}(2014){Wang}, {Zhang}, {Testi}, {Tak}, {Wu}, {Zhang},
  {Pillai}, {Wyrowski}, {Carey}, {Ragan}, \& {Henning}}]{2014MNRAS.439.3275W}
{Wang}, K., {Zhang}, Q., {Testi}, L., {et~al.} 2014, \mnras, 439, 3275

\bibitem[{{Xie} {et~al.}(1996){Xie}, {Mundy}, {Vogel}, \&
  {Hofner}}]{1996ApJ...473L.131X}
{Xie}, T., {Mundy}, L.~G., {Vogel}, S.~N., \& {Hofner}, P. 1996, \apjl, 473,
  L131

\end{thebibliography}
\end{document}